\documentclass{ws-p9-75x6-50}

\newcommand{\F}{{\cal F}}
\newcommand{\ep}{\varepsilon}
\newcommand{\Tr}{ \mbox{Tr}\; }

\newcommand{\equal}[1]
{
\buildrel{#1}\over=
}

\newcommand{\dede}[2]
{
{\partial{#1}\over\partial{#2}}
}
\newcommand{\dedi}[2]
{
{\mbox{d}{#1}\over \mbox{d}{#2}}
}
\newcommand{\formulonaX}[1]
{
  \begin{equation} 
  \begin{array}{rcl}
    #1 
  \end{array}
  \end{array}
}
\newcommand{\formulona}[2]
{
  \begin{equation}
  \begin{array}{rcl}
    #2
  \end{array}
  \end{equation} \label{#1}
}
\newcommand{\formulaX}[1]
{  \begin{equation} #1 \end{equation} 
}
\newcommand{\formula}[2]
  { \begin{equation} \label{#1} #2 \end{equation} }
\newcommand{\rif}[1]
  {(\ref{#1})}

\begin{document}

\title{{\bf Exact Renormalization Group 
in Algebraic Noncovariant Gauges\footnote{Talk given
at the Second Conference on the Exact Renormalization Group, Rome, 
September 18-22, 2000.}}}

\author{M. Simionato}

\address{ LPTHE, Universit\'e Pierre et Marie Curie 
(Paris VI) et Denis Diderot (Paris VII),\\ Tour 16, 1er \'etage,
4, Place Jussieu, 75252 Paris, Cedex 05, FRANCE\\
and\\ 
Istituto Nazionale di Fisica Nucleare, Rome, ITALY\\
E-mail: micheles@lpthe.jussieu.fr}


\maketitle

\abstracts{
I study a class of Wilsonian formulations
of non-Abelian gauge theories in algebraic noncovariant gauges
where the Wilsonian infrared cutoff $\Lambda$ is inserted as a mass 
term for the propagating fields. In this way the Ward-Takahashi identities 
are preserved to all scales. Nevertheless the BRS-invariance
in broken and the theory is gauge-dependent and unphysical at $\Lambda\neq0$.
Then I discuss the infrared limit $\Lambda\to0$. I show 
that the singularities of the axial gauge choice are avoided
in planar gauge and in light-cone gauge. Finally the rectangular Wilson loop 
of size $2L\times 2T$ is evaluated at lowest order in perturbation theory and
a noncommutativity between the limits $\Lambda\to0$ and $T\to\infty$ is
pointed out.}

\section{Introduction and conclusion}

The aim of this talk is to present in a few words the results of Ref. 1,
concerning the consistency of the Exact Renormalization Group Approach
of Wilson\cite{Wilson} in its
application to gauge theories. In particular we refer here to its most
recent formulation\cite{1PI.RG} in which the Wilsonian cutoff
$\Lambda$ is interpreted as an infrared regulator and the evolution of the
1PI effective action $\Gamma(\Phi,\Lambda)$ is studied from the
ultraviolet scale $\Lambda=\Lambda_0$ down to the physical scale $\Lambda=0$.
The application of this framework to gauge theories is 
problematic since there is no way of 
inserting the infrared cutoff consistently with gauge-invariance
(we mean with BRS-invariance): as a consequence the $\Lambda\neq0$ theory
is gauge-dependent and
unphysical and the limit $\Lambda\to0$ has to be taken strictly.

In particular one can ask three 
simple questions: i) is the $\Lambda\to0$ limit of the 1PI Green functions 
regular~? ii) is the $\Lambda\to0$ limit of the would be
physical quantities regular~? iii) 
in the computation of the interquark potential
from a Wilson loop of size $2L\times 2T$ with $T\gg L$, is the limit
$T\to\infty$ commuting with the limit $\Lambda\to0$~?

These questions can be fully answered in perturbation theory
with the following results:
i) the regularity of the $\Lambda\to0$ limit strongly depends on the 
gauge choice; in particular in the axial gauge it is not possible to
define the off-shell Green functions perturbatively,
even at one-loop level and for an Abelian theory;
ii) the $\Lambda\to0$ limit is delicate even for would-be physical
quantities like the Wilson loop: in particular the limit
does not exist in the axial gauge at order $O(g^4)$ in
perturbation theory, as proved by A. Panza and
R. Soldati (these proceedings and\cite{Soldati.Panza}); 
iii) the definition of the interquark
potential is a rather subtle question and in general the limits
$T\to\infty$ and $\Lambda\to0$ {\it do not} commute.
In the planar gauge case this is manifest even at $O(g^2)$ 
in perturbation theory. 

In general gauges, including the usual covariant
gauges, the  noncommutativity of the limits 
is expected to hold at order $O(g^4)$ in 
perturbation theory, connected with a non-exponentiation of the Wilson 
loop at $\Lambda\neq0$. We plan to elucidate this question in a future
work. The rest of this talk is devoted to a discussion of the results
i)-ii)-iii).

\section{Massive linear gauges}

As it is well known, in order to define a perturbative quantum 
field theory from a classical 
gauge invariant field theory one is forced to break gauge invariance 
through the addition of a gauge-fixing term to the action.
This procedure replaces the original local gauge symmetry with a
new global graded symmetry, the celebrated 
BRS symmetry.
The choice of the gauge-fixing term is at large extent arbitrary; however, in
order to have independence of the gauge-invariant observables from the
gauge choice, the gauge-fixing term, depending on gauge-fields, ghost
fields and auxiliary fields, must be a BRS-cocycle.
However, when an infrared regulator inconsistent with 
the BRS-symmetry is introduced, then the would be gauge-invariant 
observables become gauge-dependent. The natural expectation is that
the gauge-dependent terms are vanishing as the infrared regulator is removed.
However, this is a quite delicate point since, even if the recovering 
of the gauge
symmetry at $\Lambda=0$ has been proven for proper vertices at non-exceptional
configurations of momenta\cite{BDM.IR}, there are no theorems guaranteeing the
infrared safety of important physical quantities as the interquark potential,
which is the first thing one would like to compute in the Wilson
renormalization group approach\cite{Ellw.pot}. This motivate our
detailed study.
For concreteness here we restrict to linear gauge-fixings of kind 
\formula{lin.gauge}
{\F(A)=L^\mu(p) A_\mu(p),\quad L_\mu(p)=a n_\mu+b p_\mu\;,
}
where $a$ and $b$ are numerical parameters and $n^\mu$ is a fixed
four-vector in the Minkowsky space.
With the notations of\cite{ANC} the gauge-fixed action reads
\formula{S.BRS}
{S_\F=\int_x-\frac14F_{\mu\nu}\cdot F^{\mu\nu}+
\lambda\cdot\F(A)-\bar C\cdot\dede\F{A_\mu} D_\mu C+
\frac1{2}{\xi_2} \lambda\cdot \lambda\;,
}
or equivalently, after elimination of the auxiliary fields 
$\lambda^a$,
\formula{S.linear}
{S^{red}_{lin.g.}=\int_x-\frac14F_{\mu\nu}\cdot F^{\mu\nu}
-\bar C\cdot L^\mu D_\mu C+\frac1{2\xi_2}L^\mu A_\mu\cdot L^\nu A_\nu\;.
}
To this reduced action, we add an infrared cutoff
as a mass-like term\cite{ANC}
\formula{Gamma.0.W}
{S_{lin.g.}^{red}(A,C,\bar C,\Lambda)=S^{red}_{lin.g.}(A,C,\bar C)
+\int_x\frac12\Lambda^2A^2\;.
}
Now we can define an invertible massive propagator
for the gauge fields.
The explicit form of the propagator in 
Euclidean space and in the $\xi_2\to0$ limit
is\cite{ANC}
\formulona{D.linear}
{D_{\Lambda,\mu\nu}(p_E)
\buildrel{\xi_2 \to 0}\over=&&\frac1{ p_E^2+\Lambda^2}\delta_{\mu\nu}-
\frac{p_E\cdot L(L_\mu p_{E,\nu}+L_\nu
p_{E,\mu})}{(p_E^2+\Lambda^2)((p_E\cdot L)^2+L^2\Lambda^2)}+\cr
&&\frac{p_{E,\mu} p_{E,\nu}\ L^2}{(p_E^2+\Lambda^2)((p_E\cdot L)^2+
L^2\Lambda^2)}-\frac{\Lambda^2 L_\mu L_\nu}
{(p_E^2+\Lambda^2)((p_E\cdot L)^2+L^2\Lambda^2)}.
}
We notice that in limit $\xi_2\to0$
the propagator is both transverse
\formula{trans}
{L^\mu D_{\Lambda,\mu\nu}(p_E)\buildrel{\xi_2\to 0}\over=0,\quad\forall
\ \Lambda
} 
and invariant up to rescaling of the
gauge-fixing $L(p_E)\to C L(p_E)$.
In particular for $L_\mu\propto n_{E,\mu}$, with $n_E^2>0$,
we find the massive axial gauge whereas for light-like vectors $L_\mu=
n_{E,\mu}$ such as $n_E^2=0$ we find the massive light-cone gauge.
The characteristic feature of these gauges is that the non-Abelian
gauge symmetry can be expressed in terms of Abelian-like
Ward-Takahashi identities, which hold for any $\Lambda$.\cite{paper.I.II} 
This is the advantage with respect to the usual Wilsonian
formulation in covariant gauges with a generic cutoff function; in this  
latter case there is a highly nontrivial fine-tuning problem to solve
in order to implement correctly the (modified) Slavnov-Taylor
identities\cite{fine-tuning}.

In the axial gauge case for $n_E^\mu=(0,0,1,0)$ the propagator contains
factors $1/(p_3^2+\Lambda^2)$ which generate spurious divergences in the 
physical limit $\Lambda\to0$; these divergences which will be discussed 
in the next section. 
In the light-cone case the gauge vector has
the form $n_E^\mu=(0,0,1,i)$ and there are spurious poles of kind
$1/(p_E\cdot n_E)$ for any $\Lambda$. However,
after Wick rotation in Minkowsky space $p_0=-ip_4$, we
see that these poles are regularized with the well known Mandelstam-Leibbrandt
(ML) prescription\cite{Bassetto} which has very good analytical properties.
Therefore the light-cone gauge perturbative expansion is expected 
to be safe to all orders.
To be explicit, the Minkowskian
propagator reads
\formula{D.light-cone}
{-D_{\Lambda,\mu\nu}(p)=\frac1{ p^2-\Lambda^2+i\ep}\left\{
g_{\mu\nu}-\frac{n_\mu p_\nu+n_\nu
p_\mu}{[[p\cdot n]]}+\frac{\Lambda^2 n_\mu n_\nu}
{[[p\cdot n]]^2}\right\},
}
the spurious poles are regularized with the prescription
\formula{ML}
{\frac1{[[p\cdot n]]}\equiv\frac{p\cdot n^*}
{(p\cdot n) (p\cdot n^*)+i\ep},\quad
n=(1,0,0,1),\quad n^*=(-1,0,0,1)
}
and the propagator reduces to the standard one at $\Lambda\to0$
{\it when} the term proportianal to $n_\mu n_\nu$ can be
neglected. We point out that this is not always possible and that 
certainly the $n_\mu n_\nu$ term
cannot be neglected in the computation of infrared singular
quantities\cite{ANC}.

Finally, there is another algebraic noncovariant
gauges which has been studied in the literature, the
planar gauge\cite{Bassetto}. This gauge choice is not very convenient since 
the effective action does not satisfy simple
Ward-Takahashi identities (the reason being the presence of derivatives in the 
gauge-fixing term). Nevertheless the computation of one-loop Feynman diagrams 
in planar gauge is simpler than in light-cone gauge and its study is 
useful for sake of comparison with the other gauge choices.

We found\cite{ANC}
that the more convenient way to insert the infrared cutoff in order to have
a simple propagator is to modify the massless BRS action as
\formulona{S.planar.Lambda}
{S_{planar}(\Lambda)=&&\int_x-\frac14F_{\mu\nu}\cdot F^{\mu\nu}
-\bar C\cdot n^\mu D_\mu C+\lambda\cdot n^\mu A_\mu
\\
&&+\frac12\Lambda^2 A_\mu\cdot A^\mu+\frac12
\lambda\cdot\frac{n^2}{\partial^2+\Lambda^2}\lambda\;.
}
This gives as reduced action
\formulona{Gamma.planar}
{S^{red}_{planar}(\Lambda)=&&
\int_x-\frac14F_{\mu\nu}\cdot F^{\mu\nu}+\frac12\Lambda^2
A\cdot A+\\
&&-\frac1{2n^2}n_\mu A^\mu\cdot(\partial^2+\Lambda^2)n_\mu A^\mu-
\bar C\cdot n^\mu D_\mu C\;.
}
Notice that the ``mass'' $\Lambda^2$ multiplies the term
$$
\frac12 A^\mu \left(g_{\mu\nu}-\frac{n_\mu n_\nu}{n^2}\right) A^\nu=
\frac12 A^ig_{ij}A^j,\quad i,j\in\{0,1,2\}
$$
and thus only transverse (with respect to $n_\mu$) degrees of freedom
are screened. There are other possible ways of introducing the infrared
cutoff in the planar gauge, but this is 
the more interesting one in the sense that
one obtains a propagator which is identical 
to the propagator of the light-cone gauge, except for the fact that
the gauge vector $n^\mu$ is space-like ($n^2=-n_E^2=-1$).
As a consequence, there are spurious divergences
at $p\cdot n=0$ for any $\Lambda$. Therefore we need an explicit
prescription to manage them. The simplest choice is the CPV
prescription
\formula{CPV}
{\frac1{[p\cdot n]}\equiv\lim_{\ep\to0}\frac{p\cdot n}
{(p\cdot n)^2+\ep^2}\;.
}
We remind that in higher order computations the CPV prescription could
be problematic\cite{ANC}; however this issue is not relevant for the
one-loop computations discussed here.
  
\section{Problems of the pure axial gauge}

The standard massless version of the pure axial gauge
where the spurious divergences are managed with the CPV presciption
is affected by many serious problems\cite{Soldati} (for a recent attempt
of solving these problems in a nonstandard version of the axial
gauge see for instance \cite{Joglekar}); in particular
it is impossible to define a renormalizable perturbation theory
to all orders. The situation is similar in the Wilsonian formulation, 
at least at the perturbative
level\footnote{However, it should be noticed that the problems of
the perturbative expansion of Green functions are absent in other kinds 
of expansions of the evolution equation, like for instance
the expansion in covariant derivatives. As a matter of fact, the axial gauge
Exact Renormalization Group equation is well
defined {\em before} the expansion in powers of
the gauge fields, as stressed by J. Pawlowski at this conference. Therefore 
the axial gauge formulation can be useful for nonperturbative applications.}.

We found two major infrared problems:
i) the Fourier transform of the propagator is divergent at $\Lambda\to0$;
ii) one loop Feynman diagrams which are infrared finite in covariant gauges
becomes divergent in axial gauge for {\it all} configurations of momenta.
Moreover, even 
for would be gauge-invariant quantities as finite size Wilson loops, 
the limit $\Lambda\to0$ does not exist at order $O(g^4)$ in
perturbation theory\cite{Soldati.Panza}, 
therefore the axial gauge is definitely ruled out
for perturbative applications.
The origin of all problems comes from 
the part of the propagator proportional to $p_\mu p_\nu$,
\formula{D.pp}
{D_{\Lambda,\mu\nu}^{pp}=\frac{p_\mu p_\nu\ n^2}{(p^2+\Lambda^2)
((p\cdot n)^2+n^2\Lambda^2)}\;.
}
This quantity is a messy source of infrared divergencies due to the
identity (in the sense of distributions)
\formula{fund}
{\frac1{p_3^2+\Lambda^2}\equal{\Lambda\to0}
\frac\pi\Lambda\delta(p_3)
}
This means that for any regular function $f(p_3,\Lambda)$ such as $f(0,0)\neq0$
we have
\formula{fund2}
{\lim_{\Lambda\to0}\int_{p_3}\frac{f(p_3,\Lambda)}{p_3^2+\Lambda^2}=
\lim_{\Lambda\to0}
\frac{f(0,0)}{2\Lambda}=\infty\;.
}
The problems relative to \rif{fund} are discussed at lenght in\cite{ANC}.
Here we simply notice that in planar and light-cone gauges the double pole 
is absent and therefore these problems are avoided. 
In particular it is possible to prove that the $x-$space
propagator is regular in both gauges and that the one-loop Feynman diagrams
are well defined and with a regular limit for off-shell
configurations of momenta. At higher loops, the 
situation for the planar gauge is dubious, 
whereas the light-cone gauge is expected to be both infrared
safe for non-exceptional configurations of momenta 
and renormalizable to all orders\cite{Bassetto}.

\section{The Wilson loop test}

The Wilson loop is the simplest physical quantity 
where the effects and the problems of the infrared regularization can be 
studied. In particular our scope here is to study the subtilities of the
$\Lambda\to0$ limit and to test how the essential
property of the gauge-independence of
the Wilson loop is recovered when the infrared cutoff is removed.
In concrete in this section we compute the Wilson up to order $O(g^2)$ 
in perturbation theory. 
This is enough for elucidating various important features of  
axial, planar and light-cone gauges and it is a first step versus a more
comprehensive computation at order $O(g^4)$ in perturbation theory.

For definiteness, we shall consider a rectangular Wilson loop 
$\Gamma_{LT}$  of size $2L\times 2T$, with $T\gg L$,
\formula{rect.WL}
{W_{\Gamma_{LT}}=\frac1{N_c}
<\Tr P\exp\left(ig\int_{\Gamma_{LT}} A_\mu dx^\mu\right)>}
where $P$ denotes the path ordering on the loop
$\Gamma_{LT}$, $\Tr$ is the trace in
the fundamental representation of $SU(N_c)$ and the average is evaluated via 
a perturbative expansion of the Euclidean functional integral. 
The Wilson loop can easily be computed at order $O(g^2)$ in perturbation
theory and is given by the explicit expression
\formula{WL.2}
{W_{\Gamma_{LT}}^{(2)}=-g^2C_F\int_{-1}^1 ds_1\int_{-1}^1 
ds_2\dedi{\Gamma^\mu}{s_1}
\dedi{\Gamma^\nu}{s_2}D_{\mu\nu}(\Gamma(s_1)-\Gamma(s_2))\;,
}
with $C_F=(N_c^2-1)/(2N_c)$.
This formula can be further simplified in the $T\to\infty$ limit since
various contributions are subleading.
Moreover, due to the identity (which holds since
$\Gamma_{LT}$ is a closed path)
\formula{transv.WL}
{\int_{\Gamma_{LT}}dx^\mu dy^\nu D_{\mu\nu}(x-y)=
\int_{\Gamma_{LT}}dx^\mu dy^\nu D^T_{\mu\nu}(x-y),
}
actually only the {\it transverse} part of the propagator contributes.
At zero mass $D_{\mu\nu}^T(p)$ is the same in all gauges and this is 
the reason why the final result is gauge-independent; however at
$\Lambda\neq0$ there is a dependence on the gauge vector $n_\mu$ coming
from the part of the propagator proportional to $\Lambda^2$. 
It is this part which is problematic, in the
sense that there could be a problem of commutativity between
the limits $\Lambda\to0$ and $T\to\infty$.
This happens in the planar gauge case, where we have
a gauge-dependent contribution to the interquark potential of the kind
\formulaX
{\frac{2g^2 N_cC_F}3 \Lambda^2T^2\;
\frac{\exp(-\Lambda\cdot 2L)}{4\pi\cdot 2L}\;.
}
We see that the gauge-dependent contribution is quadratically
vanishing in the limit $\Lambda\to0$,
at finite $T$, but quadratically {\it divergent} for $T\to\infty$ at
finite $\Lambda$. The $T\to\infty$ limit {\it cannot} be taken before
the $\Lambda\to0$ limit.
This is the crucial point of our analysis. We expect this feature
become manifest for any gauge at higher orders in perturbation theory.

\section*{Acknowledgements}

I thanks the organizers and LPTHE for financial support.

\end{document}